\documentclass[conference]{IEEEtran}
\IEEEoverridecommandlockouts
\usepackage{cite}
\usepackage{amsmath,amssymb,amsfonts}
\usepackage{algorithmic}
\usepackage{graphicx}
\usepackage{textcomp}
\usepackage{xcolor}
\usepackage{soul}
\usepackage{tabularx}
\usepackage{datatool}
\usepackage[utf8]{inputenc}
\usepackage{listings}
\usepackage{csquotes}
\usepackage[T1]{fontenc}
\usepackage[most]{tcolorbox}
\definecolor{bg}{RGB}{245,248,248}
\usepackage[]{mdframed}

\def\BibTeX{{\rm B\kern-.05em{\sc i\kern-.025em b}\kern-.08em
    T\kern-.1667em\lower.7ex\hbox{E}\kern-.125emX}}

\begin{document}

%%%%%%%%%%%%%%%%%%%%%%%%%%%%%%%
%%%%%%%%%%%%%%%%%%%%%%%%%%%%%%%
%%%%%%%%%%  for arxiv  %%%%%%%%
%%%%%%%%%%%%%%%%%%%%%%%%%%%%%%%
%%%%%%%%%%%%%%%%%%%%%%%%%%%%%%%

\thispagestyle{empty}

\begin{huge}
IEEE Copyright Notice
\end{huge}

\vspace{5mm} %5mm vertical space

\vspace{5mm} %5mm vertical space

\begin{large}
© 2024 IEEE.  Personal use of this material is permitted.  Permission from IEEE must be obtained for all other uses, in any current or future media, including reprinting/republishing this material for advertising or promotional purposes, creating new collective works, for resale or redistribution to servers or lists, or reuse of any copyrighted component of this work in other works.
\end{large}

\vspace{5mm} %5mm vertical space

\begin{large}
\textbf{Submitted to:} IEEE SoutheastCon 2024 · March 15-17, 2024 -  Westin Peachtree Plaza, Atlanta, Georgia
https://ieeesoutheastcon.org/ 
\end{large}

\vspace{5mm} %5mm vertical space
Preprint Version, February 21, 2024

\vspace{5mm} %5mm vertical space

%%%%%%%%%%%%%%%%%%%%%%%%%%%
%%%%%%%%%%%%%%%%%%%%%%%%%%%
%%%%%%%%%%  paper  %%%%%%%%
%%%%%%%%%%%%%%%%%%%%%%%%%%%
%%%%%%%%%%%%%%%%%%%%%%%%%%%

\newcolumntype{L}[1]{>{\raggedright\arraybackslash}p{#1}}
\newcolumntype{C}[1]{>{\centering\arraybackslash}p{#1}}
\newcolumntype{R}[1]{>{\raggedleft\arraybackslash}p{#1}}

\title{The Pulse of Fileless Cryptojacking Attacks: Malicious PowerShell Scripts}
%\title{Fileless and Cryptojacking Hybrid Malware Attacks: A Review}

% A Review on Dangerous Combo: Fileless + Cryptojacking Malware

\makeatletter
\newcommand{\linebreakand}{%
  \end{@IEEEauthorhalign}
  \hfill\mbox{}\par
  \mbox{}\hfill\begin{@IEEEauthorhalign}
}
\makeatother
\author{

\IEEEauthorblockN{Said Varlioglu, Nelly Elsayed, Eva Ruhsar Varlioglu*, Murat Ozer, Zag ElSayed}
\IEEEauthorblockA{
\textit{School of Information Technology} \\
\textit{*School of Criminal Justice} \\
\textit{University of Cincinnati}\\
Cincinnati, Ohio, USA \\
varlioms@mail.uc.edu, elsayeny@ucmail.uc.edu, varliorr@mail.uc.edu, ozermm@ucmail.uc.edu, elsayezs@ucmail.uc.edu}
}

%\IEEEoverridecommandlockouts
%\IEEEpubid{\makebox[\columnwidth]{
%978-1-6654-0652-9/22/\$31.00 ~\copyright2024 IEEE \hfill} %\hspace{\columnsep}\makebox[\columnwidth]{ }}

\maketitle

\begin{abstract}

Fileless malware predominantly relies on PowerShell scripts, leveraging the native capabilities of Windows systems to execute stealthy attacks that leave no traces on the victim's system. The effectiveness of the fileless method lies in its ability to remain operational on victim endpoints through memory execution, even if the attacks are detected, and the original malicious scripts are removed. Threat actors have increasingly utilized this technique, particularly since 2017, to conduct cryptojacking attacks. With the emergence of new Remote Code Execution (RCE) vulnerabilities in ubiquitous libraries, widespread cryptocurrency mining attacks have become prevalent, often employing fileless techniques. This paper provides a comprehensive analysis of PowerShell scripts of fileless cryptojacking, dissecting the common malicious patterns based on the MITRE ATT\&CK framework.

\end{abstract}

\begin{IEEEkeywords}
Powershell, malicious, scripts, fileless malware, cryptojacking, cryptomining
\end{IEEEkeywords}
 
\section{Introduction}

PowerShell, developed by Microsoft for Windows, is a robust command-line shell and scripting language based on the .NET framework. It derives its strength from a rich set of cmdlets, object-oriented data handling, and seamless integration with Microsoft technologies. Widely used for system administration, automation, and scripting in Windows environments, PowerShell provides comprehensive access to critical Windows system functions, such as Windows Management Instrumentation (WMI) and Component Object Model (COM) objects. These capabilities enable remote content retrieval, in-memory command execution, and access to local registry keys and scheduled tasks \cite{hendler2020amsi}. 

PowerShell's versatility and ubiquitousness minimize the need for adversaries to customize payloads or download overtly malicious tools on a target system. Adversaries can abuse PowerShell to reflectively (filelessly) load and execute commands/executables through the invocation feature without creating new system processes \cite{RedCanary2023, afreen2020analysis}. 

Fileless malware primarily relies on PowerShell due to its robust functionality. This trend has emerged as an urgent and challenging problem \cite{hendler2020amsi} since 2017 \cite{varlioglu2022dangerous}, especially with the prevalence of common attacks such as ransomware and cryptojacking utilizing this technique \cite{olaimat2021ransomware}. 

In particular, threat actors have found this technique effective because even when the attacks are detected and the original malicious scripts are identified and removed, the processes may remain operational on victim endpoints \cite{varlioglu2022dangerous}. 

Additionally, threat actors use various open-source frameworks, including PowerShell Empire and PowerSploit, that have been created to support the use of PowerShell scripting in post-exploitation cyber-offensive activities \cite{hendler2020amsi, 8659361, 9092030}.

Threat actors have combined PowerShell-based fileless attacks with cryptojacking \cite{varlioglu2022dangerous}. Furthermore, with the emergence of new Remote Code Execution (RCE) vulnerabilities such as Log4Shell (CVE-2021-44228) affecting numerous software products across various industries, widespread fileless cryptojacking and ransomware attacks have become prevalent \cite{SzappanosGallagher2022}.

In our work, this paper has two primary contributions. Firstly, we analyzed malicious PowerShell scripts containing fileless cryptojacking scripts as a descriptive study and aimed at gaining an understanding of the problem and the overall characteristics of fileless activities. Secondly, we collected scripts providing a new dataset for this unique research area, presenting a potential avenue for further research.

\section{PowerShell-based Fileless Malware}

PowerShell-based fileless malware is considered an advanced volatile attack, as it operates entirely in system memory (RAM) without leaving an executable file on the disk for inspection \cite{bulazel2017survey}. However, in some instances, fileless malware can exhibit indirect file activity while establishing persistence by configuring a WMI (Windows Management Instrumentation) filter, allowing them to install a PowerShell command within the WMI repository. Although this approach theoretically involves a malicious WMI object residing on the disk, it does not interact with the conventional file system, making it challenging to detect and mitigate \cite{Microsoft82021}.

In fileless attacks such as Cobalt Strike, the scripts have been observed employing obfuscation techniques such as Base64-encoded commands with GunZip compression and encrypted XOR keys. That conceals the specific Windows API calls made by the shellcode and process injection into the legitimate processes. Additionally, it masks the establishment of Command and Control (C2) connections. Those techniques help threat actors hide Windows API calls and secure session data with encryption \cite{Newton2020}.

\section{PowerShell-based Fileless Cryptojacking Malware}

In-memory-only cryptojacking can run on memory exploiting PowerShell for execution \cite{handaya2020machine}. It is more dangerous than in-browser and in-host cryptojacking attacks because its evasion and persistent techniques are more sophisticated \cite{handaya2020machine}. Since fileless threats can give attackers command and control abilities with backdoors \cite{moussaileb2021survey}, a fileless cryptojacking can be converted to a data exfiltration activity or a ransomware attack. 

Similar to the fileless ransomware attacks \cite{moussaileb2021survey}, cryptojackers use open-source attack frameworks (Phase 1) to deliver malicious scripts using phishing emails or vulnerability exploitations (Phase 2) and exploit PowerShell to execute the payload on memory (Phase 3) and to create scheduled tasks for persistent mechanism (Phase 4) with continues download processes of malicious scripts\cite{sophos2019c}. Also, those malicious scripts are frequently stored and downloaded from text storage web services such as Pastebin.\cite{Gundert}. Attackers want a malicious connection to remain to spread throughout the network by escalating privileges (Phase 5) and exploiting common vulnerabilities such as the use of EternalBlue SMB vulnerability \cite{nakashima2017nsa}  or RDP brute-forcing (Phase 6). This provides the cryptojackers large pools of CPU resources (Phase 7) in victim enterprises for efficient cryptocurrency mining slaves (Phase 8) \cite{sophos2019c} to gain illicit profit with cryptocurrencies (Phase 9). A fileless cryptojacking attack can be started with phishing emails, zero-day vulnerability exploitations, and hidden scripts of malicious websites~\cite{trendmicro2019b}.

After infections, the endpoints send reports of the connection status and mining activity to the attackers' Command and control servers (C2). Attackers can also run remote commands for other purposes, such as data exfiltration or ransomware. This is another dangerous face of fileless cryptojacking compared with traditional cryptojacking techniques. Lateral movement routines are observed for spreading in victim networks \cite{Powershell1}. As a fileless threat pattern, fileless cryptojacking also uses scheduled tasks or registry keys such as "Run" or "RunOnce" for malware propagation. Also, cryptojackers can store PowerShell commands under scheduled tasks and registry keys. System information is important to run cryptojacking scripts. Thus the commands can collect computer names, GUIDs, MAC addresses, OS, and timestamp information. In the final stage, victim endpoints become slaves of mining deployers or mining pools for the illicit gains of cryptojackers.

The most prevalent fileless cryptojacking malware currently found in the wild includes Lemon Duck, Purple Fox, GhostMiner, PCASTLE, Tor2Mine \cite{varlioglu2022dangerous}. Those malware strains are primarily observed in Monero cryptocurrency (XMR) mining activities. 

\textbf{Lemon Duck}, targeting Windows systems, uses PowerShell scripts and exploits vulnerabilities with an effective lateral moving capability \cite{Lemon1}. It can use Mimikatz to dump credentials, adfind.exe to scan active directories, and many other techniques such as task scheduling, registry exploitation, WMI subscriptions for persistent mechanism \cite{lemon2}.  It can quickly infiltrate networks, starting with a single infection and swiftly spreading across the entire infrastructure to turn the resources into cryptocurrency mining slaves \cite{lemon2}. Also, its malicious scripts might be seen using the term “\$Lemon\_Duck” as a variable \cite{sophos2019c}.

\textbf{Purple Fox} exploits Windows vulnerabilities with rootkits and evasion tactics using the open-source Invoke-ReflectivePEInjection tool \cite{Invoke} to reflectively (filelessly) load and execute GZIP-compressed executables into PowerShell processes running on the memory \cite{cybereason1}.

\textbf{GhostMiner} injects malicious JavaScript into web pages to mine cryptocurrency on unsuspecting visitors' computers. It has the high-level evasion techniques \cite{caldwell2018miners}. GhostMiner exploits WMI objects as a fileless threat routine for a persistent mechanism \cite{GhostMiner1}.

\textbf{PCASTLE} utilizes PowerShell and legitimate tools for hidden mining on Windows. WannaMine, leveraging EternalBlue, mimics WannaCry and employs fileless methods for stealthy mining\cite{PCASTLE1}.

\textbf{Tor2Mine} employs PowerShell to disable security, install Monero miners, and collect Windows credentials, spreading through networks by gaining admin access and deploying services or using fileless commands when needed. Its goal is to compromise entire networks, and although it has evolved with various versions, the central strategy remains consistent: exploiting software vulnerabilities to sustain a persistent mining operation on infected networks \cite{Gallagher_2021}.

\begin{table*}[h]
  \caption{Common Command Parameters of PowerShell-based Fileless Cryptojacking}
  \begin{center}
    \begin{tabular}{l l p{7cm}}
      \hline
      \textbf{Command} & \textbf{Parameter} & \textbf{Potential Exploitation} \\
      \hline
      -nop & NoProfile & Bypass loading the user's PowerShell profile. \\
      -exec or -EP & ExecutionPolicy & Modify execution policy for the session. \\
      -bypass & Bypass & Enable script execution without restrictions. \\
      -w & Window & Open PowerShell in a new window. \\
      -hidden & WindowStyle & Start PowerShell window hidden. \\
      -c & Command & Specify a command to run in the session. \\
      -IEX & Invoke-Expression & Execute a string as a command. \\
      -enc & EncodedCommand & Use a base64-encoded command. \\
      -nologo & NoLogo & Suppress PowerShell logo at startup. \\
      -noni & NonInteractive & Run PowerShell non-interactively. \\
      -f & Force & Override restrictions and warnings. \\
      -if & InputFormat & Set script input data format. \\
      -noexit & NoExit & Keep the session open after script execution. \\
      \hline
    \end{tabular}
    \label{table1}
  \end{center}
\end{table*}

In general, threat actors exploit PowerShell for fileless cryptojacking \cite{cybereason1, Gallagher_2021} to:

\begin{itemize}
    \item Execute malicious code
    \item Obfuscate malicious activity
    \item Utilize living-off-the-land techniques
    \item Inject code into memory without touching the disk
    \item Spawn additional processes
    \item Steal system credentials
    \item Remotely download and execute arbitrary code 
    \item Establish persistence mechanisms
    \item Disable security features
    \item Free up system resources for mining
    \item Disable any competing cryptojacking
    \item Re-infect other systems on the compromised network

\end{itemize}

\section{The Dataset}

Our research is based on a dataset sourced from cybersecurity blogs and the research body. This dataset, publicly available on GitHub, comprises 200 fileless-based malicious PowerShell scripts used for cryptojacking. It offers a comprehensive collection of malicious scripts associated with well-known cryptojacking malware, including Purple Fox, Lemon Duck, Tor2Mine, and others. These scripts serve as a valuable resource for understanding cryptojacking activities, aligning with the various attack phases defined in the MITRE ATT\&CK framework \cite{Mitre2021Command}, spanning Initial Access (TA0001) to Command and Control (TA0011).

%For readers interested in accessing the dataset, you can find it on GitHub.

\section{Data Analysis with MITRE ATT\&CK Framework}

Examining the cryptojacking scripts in the context of the MITRE ATT\&CK framework \cite{Mitre2021Command} sheds light on their operational methodologies during the execution phase. This analysis deepens the understanding of cryptojacking activities and has significant implications for cybersecurity efforts and the broader research community.

In the next sections, we will explore how these scripts operate throughout the different phases below, where applicable, within the MITRE ATT\&CK framework \cite{Mitre2021Command}. This comprehensive analysis aims to provide a holistic view of the behavior and tactics of fileless cryptojacking.

\begin{itemize}
    \item TA0001 Initial Access
    \item TA0002 Execution
    \item TA0003 Persistence
    \item TA0005 Defense Evasion
    \item TA0006 Credential Access
    \item TA0007 Discovery
    \item TA0008 Lateral Movement
    \item TA0011 Command and Control
\end{itemize}

\subsection{TA0001 Initial Access}

A fileless cryptojacking attack can be started with phishing emails, exploiting public-facing applications, and hidden scripts of malicious websites~\cite{trendmicro2019b}. Especially with the emergence of new Remote Code Execution (RCE) vulnerabilities such as Log4Shell (CVE-2021-44228) \cite{CISA_2022}, and effective penetration testing tools such as Cobalt Strike, fileless cryptojacking has become effective.

In a sample Log4Shell exploitation \cite{Trellix_2022}, HTTP requests with JNDI injection payloads targeted the server with the Base64-encoded payload involving downloading and executing a script from a specified URL. This technique exploits vulnerabilities within a Tomcat server to establish unauthorized access and execute arbitrary commands on the compromised system. The attack deployed XMRig Monero miners with a PowerShell downloader script retrieving content from a Pastebin page. The other PowerShell command is associated with downloading a batch file, and subsequently executing it. It reflectively loads a Windows binary containing an encrypted, Base64-encoded loader. The batch file contains a code fragment for miner setup, specifying a wallet address. \cite{SzappanosGallagher2022}.

The forensics research reports show that threat actors can compromise a domain controller in an organization network environment, execute cryptojacking scripts, and deploy coinminers on the domain endpoints within two hours \cite{DFIR_Report_2021b}.

\subsection{TA0002 Execution}

Upon compromising an endpoint within an environment, cryptojacking scripts leverage PowerShell to execute their malicious agenda.  The scripts were seen to adapt to system variables, disable security measures, and maintain persistence while evading detection.

PowerShell scripts might offer distinct advantages to threat actors seeking to deploy a coinminer efficiently. In some sample scripts \cite{LEIN_2021}, it was observed that the actors exploit PowerShell by instantiating a \textit{"System.Net.WebClient"} object, a class designed for web interactions. The scripts then employ the class's \textit{"DownloadFile"} method to retrieve coinminer batch files from external sources, specifically GitHub repositories or Pastebin pages. This shows the maintaining agility in payload updates and modifications without altering the core script. The scripts leverage the flexibility and parameterization capabilities of PowerShell, using the \textit{"GetTempFileName"} method from the \textit{"System.IO.Path"} class to generate a temporary file for storing the downloaded batch file. Including a specific argument during the execution of the batch file allows for dynamic configuration, showcasing the threat actor's intent to customize the coinminer's behavior without modifying the script itself. Using PowerShell's living-off-the-land capabilities, such as obfuscation and temporary file usage, enhances the script's ability to minimize its footprint and avoid detection by security tools. Utilizing the \textit{"Remove-Item"} cmdlet, with the \textit{"-Force"} parameter, indicates the minimizing traces of the activities by removing the temporary file post-execution. Obfuscation and encoding of scripts are used to evade traditional antivirus solutions and enhance the overall stealthiness of the attack.

Also, cryptojacking scripts were seen using \textit{";"} as a command separator, allowing for the sequential execution of these operations.

With the breaking down of PowerShell commands in a sample attack \cite{SzappanosGallagher2022}, the command "\textit{powershell.exe -exec bypass -enc aQBl...}" employs the execution policy bypass ("\textit{-exec bypass}") to override any restrictions and specifies that the subsequent command is Base64-encoded ("\textit{-enc}"). The second command utilizes the "\textit{IEX}" alias for "\textit{Invoke-Expression}" to execute the expression. It creates a new instance of the "\textit{System.Net.WebClient}" class and retrieves content from a Pastebin page using the "\textit{DownloadString}" method.

The use of Pastebin website is frequently observed in fileless cryptojacking attacks. Pastebin is an online service for storing and sharing plain text, commonly used for source code or configuration data in the software development community. Users can upload text, and others can view and edit it as needed \cite{Pastebin}. 

Threat actors can exploit Pastebin not only to download malicious payloads but also as a straightforward command and control (C2) communication channel. When creating a pastie, they can later edit it. The scripts may be executed through a persistence mechanism, which retrieves the content of a specified Pastebin paste and dynamically adjusts its behavior based on the retrieved content. Pastebin-related scripts are observed with "curl" commands and might include parameters such as "-WindowStyle hidden," "-ExecutionPolicy ByPass," and "DownloadString." These parameters indicate attempts to hide script execution, bypass execution policies, and download content from the internet \cite{LEIN_2021}.

In another example, the inclusion of \textit{-NoP} and \textit{-NonI} parameters reflects operating without the burden of loading PowerShell profiles or engaging in interactive sessions. This allows for a swift and inconspicuous execution of the script. The \textit{-W Hidden} parameter, setting the window style to "Hidden," strategically conceals the PowerShell window during execution, adding a layer of obfuscation to the attack. Including \textit{-Exec Bypass} is pivotal in bypassing PowerShell's default execution policies, giving the threat actor the flexibility to execute the script regardless of the system's security configuration. Using environment variables, such as \textit{env:temp}, ensures the file is saved to the system's temporary directory, a common tactic to avoid raising suspicion. The subsequent \textit{Start-Process} cmdlet then initiates the execution of the downloaded executable.

To carry out their tasks, these scripts use various classes and methods, each serving a specific purpose. This includes classes such as "\textit{text.encoding}" for character encoding, "\textit{IO.StreamReader}" for stream reading, and "\textit{net.webrequest}" for handling web requests. Key methods involved in these operations comprise "\textit{getbytes()}" for converting data to UTF-8 encoding, "\textit{readtoend()}" to extract the entire content of a stream, and "\textit{FromBase64String()}" for decoding Base64-encoded data. Additionally, the scripts also work with classes like "\textit{RSAParameters}" and "\textit{RSACryptoServiceProvider}" for encryption tasks.

Lemon Duck scripts also leverage "\textit{certutil}" within PowerShell to download and execute malicious scripts, including compiled Python executables. These scripts can be executed using Windows schedule tasks, leading to the execution of malicious PowerShell commands and the download of a Cobalt Strike beacon. These payloads initiate a Cobalt "\textit{beacon}" within the newly spawned PowerShell process memory, aiming to establish communication with a command and control server \cite{Nataraj_2021}.

Also, cryptojacking scripts were seen by downloading the XMRig ZIP file from a remote server and installing it. XMRIG is an open-source software for mining cryptocurrencies like Monero or Bitcoin \cite{Darley_Robinson_Ellis_2022}. PowerShell command employs the "\textit{System.IO.Compression.ZipFile}" class to extract the contents of "\textit{xmrig.zip}" to "\textit{mimu6}". Mimu is a variant of the XMrig Monero miner botnet \cite{SzappanosGallagher2022}. The script downloads additional tools, such as 7za.exe and nssm.zip, if necessary. 7z is a popular file archiver (compression) tool. NSSM is also a tool that ensures if the application running as a service fails, it will be restarted. NSSM is seen as a persistence helper of the XMrig mining bot \cite{Patterson}.

The fileless part of the execution phase is seen with PowerShell commands, which modify the "config.json" file within the XMRig directory, replacing placeholders with specified values for URL, user, password, CPU usage, and log file. Conditional statements determine the startup directory and execute different setups based on the value of the "\textit{ADMIN}" variable \cite{Darley_Robinson_Ellis_2022}.

One of the other patterns was Base64-encoded commands in the cryptojacking scripts. The encoded commands were observed with specific configurations, including the suppression of profile loading (\textit{"-NoP"}), non-interactive mode (\textit{"-NonI"}), hidden window (\textit{"-W Hidden"}), and bypassing execution policies (\textit{"-Exec Bypass"}). The decoded versions of those commands show that downloader commands aim to fetch and execute additional PowerShell codes. 

%%%%%%%%%%%%%%%%%%%%%%%%%%%%%
%%%%%%%%%%%%%%%%%%%%%%%%%%%%%
%%%%%%%%%%%%%%%%%%%%%%%%%%%%%
%%%%%%%%%%%%%%%%%%%%%%%%%%%%%
%%%%%%%%%%%%%%%%%%%%%%%%%%%%%

\subsection{TA0003 Persistence}

In fileless cryptojacking attacks, threat actors have been observed establishing persistence by exploiting specific Windows components, including Task Scheduler and Windows Management Instrumentation (WMI) event subscriptions. Task Scheduler is a feature in Windows that enables the automation of various tasks at specified intervals or in response to specific events. Concurrently, Windows Management Instrumentation (WMI) is leveraged to install event filters, providers, consumers, and bindings that execute code when predefined events occur.

In a sample command, threat actors utilized the \textit{"schtasks"} command to create a new scheduled task with various parameters. The \textit{"/create"} flag indicates the intent to create a task, and the \textit{"/sc MINUTE"} parameter schedules the task to run at specified minute intervals. The \textit{"/F"} flag enforces the creation of the task, overwriting it if it already exists.

Additionally, the \textit{"/ru"} system parameter is specified to run the task under the \textit{"system"} account, a high-privileged built-in account in Windows systems. These scripts interact with the Task Scheduler service to verify the user's administrator privileges. Tasks are set up under the "\textit{system}" account; otherwise, they are created for the current user. The initial step involves checking the current Windows identity using the "\textit{GetCurrent}" method from the "\textit{System.Security.Principal.WindowsIdentity}" class. If the identity does not contain the string "\textit{SYSTEM}," the script deletes all existing scheduled tasks using "\textit{SchTasks.exe /Delete /TN * /F}".

The \textit{"/RL HIGHEST"} parameter sets the task to run with the highest privileges. The command "\textit{mshta <URL>.hta}" launches the \textit{mshta} utility to open a web page. The \textit{"-w hidden -c <VARIABLE>"} parameter defines the action executed when the task runs, launching a hidden PowerShell command with the \textit{-w} hidden flag and executing a PowerShell script. Additionally, it pauses execution for a specific period using the \textit{"start-sleep"} command. The configuration includes PowerShell commands that download and execute code from specified Pastebin URLs \cite{Darley_Robinson_Ellis_2022}.

The scripts involve the registration of event filters, consumers, and bindings, typically associated with Windows Management Instrumentation (WMI) event handling. In a sample script, a conditional block attempts to create a WMI event filter and consumer. The "\textit{Set-WmiInstance}" cmdlet configures the event filter and consumer with specified names. The event query, constructed with the "\textit{SELECT}" statement, targets "\textit{Win32\_PerfFormattedData\_PerfOS\_System}" within a 300-second timeframe. The \textit{-enc} flag executes a base64-encoded PowerShell command, hidden and non-interactive, to execute a script downloaded from Pastebin. An empty catch block is present in case of an error during WMI instance creation.

To retrieve a WMI event filter from the "\textit{root/subscription}" namespace, the scripts use the "\textit{Get-WMIObject}" command, specifying \textit{"-Class \_\_EventFilter"} and \textit{"-NameSpace root/subscription"}.

The scripts utilize the PowerShell \textit{"getRan()"} function to generate random strings for task names, ensuring uniqueness and unpredictability. COM objects representing the Windows Task Scheduler service are created, connecting to it using the \textit{"Connect()"} method. An attempt is made to retrieve a task from the root folder of the Task Scheduler service by calling \textit{"GetFolder"} and \textit{"GetTask"} methods.

These cryptojacking scripts, under scheduled tasks, target servers on the default RDP port (\textit{"3389/TCP"}), attempting access using common administrator usernames. On success, the script adjusts Windows Firewall settings to open a specific TCP port, acting as a marker. The continuous exploitation code operates with a specific minute pause, generating new, random IP addresses to compromise SMB and MS-SQL services. It collects machine profiling data for transmission to a command-and-control server, and modifying Windows Firewall settings serves as a persistence mechanism.

The scripts attempt to delete specific scheduled tasks, possibly to cover tracks or remove traces. In preparation for installing a cryptocurrency miner, they stop services, terminate processes, and remove executable files. Commands like \textit{"net stop"} and \textit{"taskkill /f"} halt services and terminate processes, while \textit{"cmd /c del"} deletes certain executable files.

In Purple Fox attacks, the scripts enter loops, checking for the existence of the registry key. These loops act as a form of persistence, ensuring continuous execution until specified registry keys are set \cite{KRISTAL_2020}.

%%%%%%%%%%%%%%%%%%%%%%%%%%%%%%%%%%%%%%%%
%%%%%%%%%%%%%%%%%%%%%%%%%%%%%%%%%%%%%%%%
%%%%%%%%%%%%%%%%%%%%%%%%%%%%%%%%%%%%%%%%
%%%%%%%%%%%%%%%%%%%%%%%%%%%%%%%%%%%%%%%%
%%%%%%%%%%%%%%%%%%%%%%%%%%%%%%%%%%%%%%%%

\subsection{TA0005 Defense Evasion}

Threat actors exhibit sophisticated evasion tactics for fileless cryptjoacking. Each subsection below provides insights into the intricate techniques utilized by threat actors to evade detection and carry out malicious activities discreetly.

\begin{itemize}
    \item Manipulating Antivirus Products
    \item Uninstalling Antivirus Products
    \item Firewall Disabling
    \item Dynamic Delay and Content Fetching
    \item Steganographic Tactics
    \item Reflective PE Injection
    \item Dynamic Payload Evasion
    \item Disrupting Competing Mining Services
\end{itemize}

\subsubsection{\textbf{Manipulating Antivirus Products}}

Fileless cryptojacking scripts were observed using a method that involved inserting exclusion rules into Windows Defender. The command, encapsulated within a try-catch block, was intended to exempt the entire C drive from virus scans utilizing the \textit{"Add-MpPreference"} cmdlet. This deliberate exclusion enabled malicious actors to download and install tools on the C drive without setting off virus scans. Furthermore, the commands featured base64-encoded payloads crafted to retrieve and execute the subsequent phase of the attack, thereby bolstering their capacity to conduct malicious activities discreetly.\cite{CISA_2022}.

Also, the scripts were seen to register custom commands to initiate trusted Windows processes, like \textit{CompMgmtLauncher.exe} or \textit{ComputerDefaults.exe}, and subsequently remove the associated registry keys to disable Windows Defender real-time monitoring and add exclusions for specific paths and processes \cite{SZELES_2020}.

\subsubsection{\textbf{Uninstalling Antivirus Products}}

Furthermore, cryptojacking scripts leverage PowerShell to uninstall various antivirus products, including ESET, Kaspersky, and Avast. This uninstallation is achieved using the Windows Management Instrumentation Command-line (WMIC) with the "\textit{"call uninstall"} command to invoke the uninstall operation on anti-malware tools.

Specifically, the Windows Management Instrumentation Command-line (\textit{"wmic.exe"}) employs the \textit{"/nointeractive"} parameter to ensure unattended removal, eliminating the need for user interaction during the uninstallation process. Additionally, the command involves executing the uninstaller with the \textit{"/verysilent"} parameter, ensuring an extremely silent uninstallation, as well as \textit{"/suppressmsgboxes"} to prevent the display of message boxes during the uninstallation. The \textit{"/norestart"} parameter avoids a system restart after removing the antivirus product.

Lemon Duck also attempts to uninstall security products from the machine through WMI, employing "\textit{taskkill}" for forced process termination and Windows service controller commands to disable and remove these security products \cite{Nataraj_2021}.

In Tor2Mine cryptojacking attacks, the scripts were observed using the \textit{"sc (Service Control)"} command to stop and delete services associated with antivirus products. This disrupts the normal operation of these security services. Also, registry modifications are made to disable Windows Defender. This involves setting \textit{"DWORD"} values in specific registry keys to \textit{"1,"} disabling of antivirus features with the Windows Defender parameters below \cite{Gallagher_2021}.
\begin{itemize}
    \item DisableAntiSpyware
    \item DisableBehaviorMonitoring
    \item DisableOnAccessProtection
    \item DisableScanOnRealtimeEnable
\end{itemize}

\subsubsection{\textbf{Firewall Disabling}}

In addition to the fileless cryptojacking scripts and the strategic exclusion from Windows Defender scans, the threat actors took further measures to enhance their attack capabilities. Specifically, they disable Windows firewall by setting all profiles to state=off
\textit{"netsh advfirewall set allprofiles state off"} \cite{Darley_Robinson_Ellis_2022}. This command uses the \textit{"netsh"} utility to modify the configuration of the Windows Advanced Firewall. In particular, it sets the state of all firewall profiles (Domain, Private, and Public) to "off," effectively turning off the firewall protection for all network profiles.

Disabling the Windows Firewall removes a significant layer of defense that the operating system provides against unauthorized network access and inbound/outbound communication. By turning off the firewall, the threat actors aim to create a more permissive environment for their malicious activities, allowing unrestricted communication and data transfer without impeding the firewall's security measures.

\subsubsection{\textbf{Dynamic Delay and Content Fetching}}

The other technique is to use the \textit{"Start-Sleep -Seconds <Integer>"} command for a deliberate delay of specified seconds to evade immediate detection and to ensure a specific state on the compromised system. Following the delay, the script utilizes the \textit{"(New-Object System.Net.WebClient).DownloadString"} method. The method is directed to fetch the contents of a remote URL. This dynamic retrieval of content from a remote server indicates a strategy threat actors employ to avoid static signatures and continuously adapt their tactics.

\subsubsection{\textbf{Steganographic Tactics \& Disguising Malicious Payloads in JPG \& PNG Files}}

In fileless cryptojacking attacks, threat actors use the "steganography" technique to conceal malicious activities within seemingly innocuous JPG or PNG image files, as seen in Fig.~\ref{fig1} \cite{KRISTAL_2020}. The image file is downloaded and then extracted from memory. The embedded encoded code is decoded, and the payload is run.

\begin{figure}
    \centering
    \vspace{0.1cm}
    \includegraphics[width=1\linewidth]{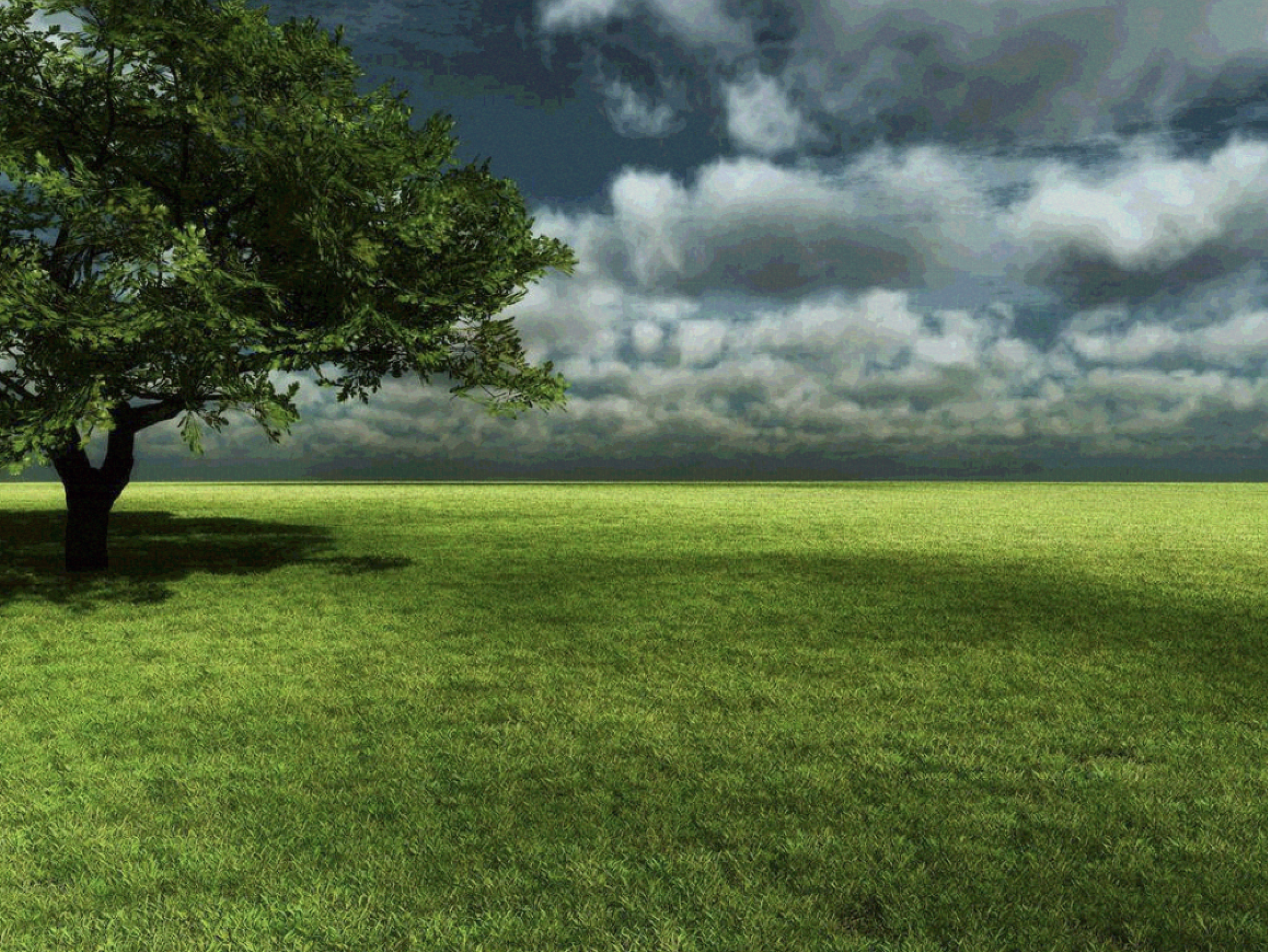}
    \caption{Sample image file for a fileless cryptojacking steganography \cite{KRISTAL_2020}.}
    \label{fig1}
\end{figure}

The embedded code below is an obfuscated PowerShell script to download and execute malicious content from a remote server. Specifically, it creates a \textit{"Bitmap"} object from an image fetched from a specified URL, then processes the pixel data to form a payload, and finally executes the payload using the IEX (Invoke-Expression) command.

\begin{tcolorbox}[colback=bg,boxrule=-4pt,arc=0pt]
{
\begin{small}
\begin{verbatim}
$uyxQcl8XomEdJUJd='sal a New-Object;Add-Type 
-A System.Drawing;$g=a System.Drawing.Bitmap 
((a Net.WebClient).OpenRead(" 
http[:]//rawcdn.githack[.]cyou/up.php?key=3")
;$o=a Byte[] 589824;(0..575)|%{foreach($x 
in(0..1023)) {$p=$g.GetPixel($x,$_)
;$o[$_*1024+$x]=([math]::Floor
(($p.B-band15)*16)-bor($p.G -band 15))}};
IEX([System.Text.Encoding]::ASCII.GetString
($o[0..589362]) IEX ($uyxQcl8XomEdJUJd)
\end{verbatim}
\end{small}
}
\end{tcolorbox}

Steganography allows threat actors to hide executable code, scripts, or other malicious payloads within the pixels or metadata of an image file without visibly altering its appearance \cite{KRISTAL_2020}. In PurpleFox scripts, it was seen that JPG files were malicious MSI installation packages. PowerShell scripts can be camouflaged as a .jpg image file. Upon execution, the script performs malicious actions, including the download and execution of Purple Fox's main component or attempts at privilege escalation \cite{foregenix2022}. In the sample attack, the PowerShell script showcases a set of actions executed if the current user possesses administrative privileges. The script dynamically defines a \textit{".NET"} type named \textit{"msi"} using the \textit{"Add-Type"} cmdlet, encapsulating native Windows Installer API functions within this custom type. Subsequently, the \textit{"MsiInstallProduct"} function specifies the URL with a jpg extension as the package path \cite{Purple1}.

The cryptojacking scripts might have URLs also specifying PNG file formats \cite{Bojan_2022}, traditionally an image format. However, in a sample script, it was seen that the commands invoke the Windows Installer (Msiexec) to install the content retrieved from the URL.  In a sample command, \textit{"Msiexec /i http://<DomainAddress>/<FileName.Png /Q"}, the \textit{"/i"} flag indicates an installation process, while the URL specifies the location of the payload disguised as a PNG image. The \textit{"/Q"} flag ensures a quiet installation without user interaction \cite{Bojan_2022}.

In the steganography technique, the scripts use \textit{"System.Drawing"} classes and manipulating pixel data. In a PowerShell script, the encoded strings are decoded and executed using the \textit{"IEX Invoke-Expression"} cmdlet. \textit{"System.Drawing"} assembly is then loaded using \textit{"Add-Type"}. Subsequently, the script downloads an image from a specified URL using \textit{"Net.WebClient"} and creates a \textit{"System.Drawing.Bitmap"} object from the downloaded image via the \textit{("New-Object Net.WebClient).OpenRead"}method. The script then iterates through each pixel of the image, manipulates the RGB components, and stores the resulting values in an array. The script concludes by executing a command obtained by decoding a portion of the data within the array using another instance of IEX \cite{foregenix2022}.

Byte arrays are used to store pixel information. The script processes each row of pixels in the image, extracting information from the specific components. The extracted data is then encoded into the byte array. The encoded payload is converted from \textit{"ASCII"} encoding using \textit{"[System.Text.Encoding]::ASCII.GetString(<Byte Array>)"}. This payload is then executed using \textit{"Invoke-Expression"} command. The intentional delays were observed with \textit{"Start-Sleep <Integer>"} that allows the script to evade immediate detection \cite{Bojan_2022}.

\subsubsection{\textbf{Reflective PE Injection}}

Threat actors frequently utilize the open-source PowerSploit framework's "Invoke-ReflectivePEInjection" module for process injection in fileless attacks. This module mimics reflective DLL injection, a technique commonly employed by adversaries \cite{cybereason1}.

Purple Fox attacks leverage reflective PE injection to evade detection mechanisms. The injected code operates entirely in memory without leaving traces on the disk. The \textit{"Invoke-ReflectivePEInjection"} function encapsulates this technique, enabling the injection of a Portable Executable (PE) binary directly into the memory space of another process. By fetching the reflective DLL and executable from remote URLs, adversaries enhance the fileless nature of their attack, making it challenging for traditional antivirus solutions to detect and mitigate. The overarching goal of such a technique is often to establish persistence, execute arbitrary commands, or deploy additional malicious payloads without relying on conventional file-based execution methods. In the sample PowerShell scripts, the invocation of the \textit{Invoke-ReflectivePEInjection} function comes with a URL \textit{"http://<domain>/<filename>.jpg"} that signifies the source of the reflective DLL, while the second URL \textit{"http://<domain>/filename.jpg"} serves as the target executable into which the reflective DLL is intended to be injected. The script executes a silent installation of an MSI package, employing the command \textit{"msiexec /i <VARIABLE> /q"}. The script embeds native Windows Installer API functions with the MSI package. This script showcases a multi-faceted attack strategy where a reflective DLL, obtained from a disguised image file, is injected into a seemingly innocuous executable. The incorporation of a silent MSI package installation further underscores the threat actor's intention to carry out malicious activities without drawing attention. The use of reflective PE injection not only enhances the fileless nature of the attack but also aims for persistence and stealth, evading conventional security measures \cite{Purple1}.

In these scripts, employing the \textit{"Invoke-ReflectivePEInjection"} cmdlet with the \textit{"-ExeArgs"} parameter provides additional arguments to the executed process, including PowerShell flags such as \textit{"-nop", "-windowstyle hidden", "-exec bypass", and "-EncodedCommand"}. These flags collectively enhance stealth and bypass PowerShell execution policies, showcasing the threat actors' focus on evasion and maintaining a low profile during exploitation. This technique was mostly seen leveraging steganography for advanced evasion techniques in fileless cryptojacking attacks.

\subsubsection{\textbf{Dynamic Payload Evasion}}

Threat actors use advanced obfuscation and runtime loading in PowerShell. Threat actors leverage such intricate obfuscation and dynamic loading techniques to bypass static security measures. In a sample script, the usage of dynamic assembly loading allows adversaries to fetch and execute payloads on-the-fly, making it challenging for traditional antivirus solutions to detect and block the malicious activities at the initial stage. The repetition of this process indicates an attempt to ensure persistence and to achieve a specific evasion goal. Within each iteration, the script downloads data from a specified URL using \textit{"Net.WebClient"} and loads it as an assembly with \textit{"[System.Reflection.Assembly]::Load"}. The specific sleep period between iterations adds a delay, enhancing the script's stealthiness and complicating detection efforts. 

Fetch and execute payloads on-the-fly" refers to a technique where a piece of code, typically malicious, retrieves additional components or instructions dynamically from an external source during runtime. In the context of cybersecurity, this often involves downloading malicious content or instructions from a remote server or source, allowing threat actors to adapt and evolve their attacks in real-time. This dynamic behavior enhances the attackers' ability to evade static security defenses, as the specific payload or instructions are not present in the initial code but are fetched during execution.

In the sample PowerShell script, the script downloads data from a remote server, loads it as an assembly, and executes the retrieved payload. This dynamic loading and execution occur on-the-fly, meaning that the actual content and actions are determined and executed during the script's runtime rather than being predefined in the static code.

\subsubsection{\textbf{Usage of VBScripts}}

Threat actors leverage VBScripts as a versatile tool for executing commands aimed at network manipulation, file system tampering, disruption of normal system operations, and manipulating Windows Management Instrumentation (WMI). 

In a sample script, it executes a series of \textit{"netsh"} commands, allowing threat actors to modify \textit{"IPv4"} and \textit{"IPv6"} settings, establish \textit{"IPsec"} policies and filter lists, and control network traffic by adding filters on specific ports and protocols. This demonstrates threat actors interact with the Windows environment, enabling the alteration of critical network configurations via VBScripts. 

Additionally, the script showcases attempts to manipulate default system files, such as \textit{"jscript.dll"} and \textit{"cscript.exe"}, by taking ownership and adjusting access permissions. The \textit{"takeown"} command is used to take ownership of the specified files, allowing threat actors to gain control over them. Subsequently, the \textit{"cacls"} command is utilized to modify access control lists (ACLs), restricting permissions for the "\textit{"everyone"}" group and effectively denying access to these files. This strategy aims to impede the execution of scripts and hinder security mechanisms that rely on these files. 

The script also uses the \textit{Start-Sleep} and \textit{"Restart-Computer -Force"} commands. The deliberate use of delays, such as the \textit{"15-minute"} pause before a system restart, suggests an evasion tactic to evade immediate detection and response. Overall, threat actors employ VBScripts due to their ease of use, integration with Windows systems, and ability to execute various commands, making them a potent tool in the arsenal of cyber adversaries for carrying out disruptive actions \cite{KRISTAL_2020}.

Furthermore, the script employs conditional statements to execute different payloads depending on the identified OS version. For instance, it attempts to install an MSI package silently when running on \textit{"Windows XP"} or \textit{"Windows Vista"}, and it executes encoded PowerShell commands when running on \textit{"Windows 7"}.

\subsubsection{\textbf{Disrupting Competing Mining Services}}

On the other hand, the scripts were observed with preparing the mining environment by removing previous miner instances. The scripts employ several built-in methods and parameters, such as \textit{"sc stop"} and \textit{"sc delete"} for stopping and deleting other miners services such as named \textit{"gado\_miner"} and \textit{"taskkill /f /t"} to forcefully terminate the processes such as \textit{"xmrig.exe"}, \textit{"logback.exe"}. The scripts attempt directory removal using \textit{"rmdir /q /s"} to recursively delete the other miner directories such as "\textit{"mimu6"}". An error-checking mechanism verifies the successful removal of the directory and repeats the removal attempt until successful or until a specified condition is met \cite{Darley_Robinson_Ellis_2022}.

The scripts also use \textit{"taskkill /F /PID <PID>"} command to eradicate processes and files associated with cryptocurrency mining activities systematically targeting specific process names. The scripts terminate them and remove their associated temporary directories in the Temp folder.

Also, the scripts include \textit{"netstat -ano | findstr TCP"} to identify any process operating on ports \textit{3333, 4444, 5555, 7777, 9000} and stop the running processes \cite{Darley_Robinson_Ellis_2022} to disable the other mining services.

\subsection{TA0006 Credential Access}

Upon compromising an endpoint within an environment, cryptojacking scripts turn to PowerShell for credential access. They use brute-force attacks on the other server account credentials, such as the Microsoft SQL server. This process involves utilizing a password, hash dictionary, and an array of NTLM hashes in a "\textit{pass the hash}" attack. This phase encompasses activities focused on acquiring valid account credentials to gain unauthorized access to systems and resources.

Threat actors commonly utilize Mimikatz by executing the commands specific parameters such as \textit{"log"} to enable logging, \textit{"privilege::debug"} to grant debug privileges, \textit{"sekurlsa::logonpasswords"} for extracting plaintext passwords from the LSASS memory, and \textit{"sekurlsa::tickets /export"} to export Kerberos tickets, all culminating with the exit directive. The attackers also employ a VBScript (launch.vbs) to execute Mimikatz with elevated privileges, requesting administrative access through runas extracting credentials and exporting Kerberos tickets \cite{DFIR_Report_2021b}.

In a compromised environment, cryptojacking scripts leverage PowerShell to randomly generate IP addresses for potential targets, enabling port scans and exploiting vulnerabilities in services like SMB and MS-SQL. To compile a list of target IP addresses, these scripts use PowerShell features like \textit{"ipconfig"} and \textit{"netstat"}, along with regular expression matching, array manipulation, and web requests.

Attackers scan for vulnerable machines, specifically on various versions of the Windows operating system. The scan identifies potentially vulnerable endpoints, and the attacker exploits the target system upon discovery. Specifically, these scripts use PowerShell's "\textit{System.Net.Sockets.TcpClient}" .NET class to establish network connections, primarily for scanning IP addresses. The scripts focus on identifying open port \textit{445} for exploitation. Depending on the Windows version, the script executes specific functions, ultimately compromising the target and advancing the attacker's goals. This phase involves active code execution to exploit vulnerabilities and gain unauthorized access. It is worth noting that Lemon Duck scripts employ the PingCastle EternalBlue vulnerability scanner \cite{PingCastle_2022}.

\subsection{TA0007 Discovery}

Upon securing a substantial foothold within the network with the executions and persistence mechanisms of cryptojacking attacks, the threat actors execute PowerShell commands targeting the Active Directory. A sample command "\textit{powershell.exe get-adcomputer -filter * -properties * | select name, operatingsystem, ipv4address >}", was seen as designed to extract a comprehensive list of all machines affiliated with the domain. This action provided the malicious actors with valuable information, including the names, operating systems, and IPv4 addresses of connected machines, enabling them to strategize further and escalate their activities within the compromised network \cite{CISA_2022}.

Also, the cryptojacking scripts were observed using the command \textit{"net accounts"} to display and configure the network-wide user account settings on a local system \cite{DFIR_Report_2021b}. 

Additionally, the scripts were seen downloading strings from remote servers, passing encoded information about the system's domain and the presence of specific software such as Veeam \cite{SzappanosGallagher2022}. The specific parameters, representing the URL-encoded domain and the specific software presence information, respectively, are appended to the URL. This dynamic parameterization allows threat actors to customize the information sent to the remote server based on the specific attributes of the target system. By querying the system's domain and identifying the presence of specific software, attackers gain insights into the organizational structure and potentially exploit vulnerabilities associated with specific software installations.

\subsection{TA0008 Lateral Movement}

Lemon Duck cryptojacking employs various PowerShell techniques for lateral movement within a compromised network. The scripts are observed utilizing EternalBlue for SMB exploitation, scanning machines for vulnerabilities, and launching attacks based on the target's Windows OS version. In a sample script, the actors scan the network that responds on 445/TCP using a tool called PingCastle to see if they are susceptible to the EternalBlue vulnerability. Additionally, the can leverage USB and network drives by writing malicious Windows *.lnk shortcut files and DLL files to removable storage connected to infected machines. The scripts might also engage in Pass-the-Hash attacks, verifying user privileges and using NTLM hashes to upload malicious scripts and associated files to remote machines in the network. Furthermore, the malicious scripts may conduct MS-SQL Server brute-forcing by attempting various passwords for the specific user accounts and RDP brute-forcing by cycling through a list of hardcoded passwords to gain unauthorized access. These lateral movement techniques, coupled with the continuous monitoring and reporting to the command-and-control server, enable Lemon Duck to persistently propagate and conduct cryptojacking activities across the network \cite{sophos2019c}.

Also, the sample scripts show that the script employs a \textit{"foreach loop"} to iterate through a list of IP addresses with open RDP ports stored in an array. For each IP address, the script checks whether it meets specific conditions, including not being present in an array and having a length greater than \textit{"6"} characters. This ensures that the script focuses on unique and potentially vulnerable RDP targets. Within the loop, the script initiates an RDP brute-force attack by iterating through a list of passwords stored in an array . For each password, it uses an object (\textit{"RDP.BRUTE"} in the sample script), instantiating it with the current IP, the username "administrator," and the password for the RDP login attempt \cite{sophos2019c}.

\subsection{TA0011 Command and Control}

Cryptojacking scripts, executed within PowerShell, were observed to have command and control (C2) capability to report important machine profile details and module statuses to a central C2 server. By collecting essential system data like MAC address, UUID, and computer name, the script utilizes the \textit{Net.WebClient }class and the \textit{DownloadString} method to relay this information. It transmits comprehensive data, encompassing system statistics, open port counts, IP addresses, and module execution details. This continuous communication allows the attacker to maintain real-time oversight of the compromised system's environment, adjusting their strategy based on insights into user accounts, system configuration, privileges, exploitation and mining module statuses, and more. The scripts also manipulate port openings and firewall rules and set up port proxies.

The script leverages the native \textit{netsh} command-line utility, employing a range of parameters and arguments for various tasks. These include the "\textit{firewall}" parameter, which configures the Windows Firewall; the "\textit{add portopening}" parameter used to create new port opening rules in the Windows Firewall; and the \textit{"TCP"} parameter specifying rules for TCP traffic. Additionally, the script utilizes "\textit{interface portprox}" to configure port proxying on the network interface. It employs "\textit{add v4tov4}" to create IPv4-to-IPv4 port proxy rules, facilitating traffic forwarding from one IPv4 address/port to another IPv4 address/port. "\textit{Listenport}" designates the port where the proxy listens for incoming traffic, while "\textit{connectaddress}" determines the destination IPv4 address to which incoming traffic is forwarded. Lastly, "\textit{connectport}" specifies the port at the destination address to which incoming traffic is forwarded.

The unauthorized cryptocurrency mining activity, in addition to cryptojacking, poses a risk of deploying extra malicious scripts for remote command and control (C2). In a sample attack \cite{SzappanosGallagher2022}, backdoors were introduced alongside the cryptojacking operation, showcasing the script's dual functionality. The script establishes a persistent reverse shell through a continuous '\textit{"while"}' loop, utilizing a TCP connection with the \textit{"Net.Sockets.TCPClient"} class to communicate with a remote server. Bidirectional communication allows threat actors to execute commands remotely, enabling activities like data exfiltration and lateral movement. The "\textit{GetStream()}" method obtains a network stream for data transmission, and \textit{"IO.StreamWriter"} facilitates writing data using the \textit{"WriteToStream"} function. The continuous loop and dynamic command execution enhance flexibility. Features like a prompt \textit{"SHELL>"} contribute to maintaining interactivity. Subsequently, the script decodes and executes received commands using \textit{"Invoke-Expression"}. The command output is returned to the attacker by establishing a bidirectional communication channel through \textit{"WriteToStream"}. The inclusion of a delay, DNS client cache clearing using \textit{"Clear-DnsClientCache"}, and hidden window style contributes to persistence, bidirectional communication, and evasion techniques, highlighting its adaptability for multifaceted malicious operations.

\section{Conclusion}

In this paper, we first reviewed PowerShell-based fileless cryptojacking attacks, specifically focusing on unauthorized coinminer deployments that operate in-memory (RAM), originating from prevalent cryptojacking malware families. Subsequently, we analyzed collected fileless cryptojacking scripts within the MITRE ATT\&CK framework. Our findings highlight an increased effectiveness in fileless cryptojacking, attributed to the rise of new Remote Code Execution (RCE) vulnerabilities and the integration of penetration testing tools like Cobalt Strike, aligning it with the prevalence of another common threat, ransomware. During the execution phase, these scripts consistently leverage standard PowerShell exploitation parameters to evade detection, often utilizing platforms like Pastebin. In the persistence phase, the employed techniques share similarities with other attacks, but a notable observation is the intentional disabling of competitive coinminers. This logic extends to the defense evasion step, wherein exclusions for coinminer scripts are deliberately established. The credential access, discovery, lateral movement, and command-and-control phases exhibit similar attack patterns consistent with attacks aiming for domain-wide impact.

The identified patterns are summarized in a table, and specific details regarding these patterns are elaborated in the research. This offers an opportunity to leverage Natural Language Processing (NLP) and deep learning methods for analyzing the textual data of PowerShell scripts, facilitating the development of more effective detection methods in future research.

\bibliographystyle{IEEEtran}
\bibliography{main}

\end{document}